\documentclass[aps,notitlepage, amsmath, showpacs, superscriptaddress, onecolumn,sort&compress,floatfix, amssymb,noeprint]{revtex4-1}

\usepackage{float}
\usepackage{graphicx}
\usepackage{amssymb}
\usepackage{color}
\usepackage{subfigure}
\usepackage{epsfig}
\usepackage[normalem]{ulem}
\usepackage{braket}
\usepackage{hyperref}

\expandafter\let\csname equation*\endcsname\relax
\expandafter\let\csname endequation*\endcsname\relax
\usepackage{amsmath}

\begin{document}

\title[Interferometric measurement of micro-g acceleration with levitated atoms]{Interferometric measurement of micro-g acceleration with levitated atoms}

\author{A. Di Carli, C. D. Colquhoun, S. Kuhr, E. Haller}

\address{University of Strathclyde, Department of Physics,
Scottish Universities Physics Alliance (SUPA), Glasgow G4 0NG, United Kingdom}

\begin{abstract}
The sensitivity of atom interferometers is usually limited by the observation time of a free falling cloud of atoms in Earth's gravitational field. Considerable efforts are currently made to increase this observation time, e.g.\ in fountain experiments, drop towers and in space. In this article, we experimentally study and discuss the use of magnetic levitation for interferometric precision measurements. We employ a Bose-Einstein condensate of cesium atoms with tuneable interaction and a Michelson interferometer scheme for the detection of micro-g acceleration. In addition, we demonstrate observation times of 1s, which are comparable to current drop-tower experiments, we study the curvature of our force field, and we observe the effects of a phase-shifting element in the interferometer paths.
\end{abstract}

\maketitle

\section{Introduction}

Precision measurements with matter waves have shown tremendous advances over the last decades. In particular, atomic matter wave interferometers demonstrated a ground-breaking increase of the measurement precision of inertial effects, such as rotation \cite{Gustavson1997, Dutta2016} and acceleration \cite{Peters1999, Hardman2016}. In addition, atomic matter wave interferometers have been used to determine the fine-structure constant \cite{Bouchendira2011}, Newton's gravitational constant \cite{Rosi2014,Prevedelli2014}, and constraints on dark energy \cite{Hamilton2015}. Similar to optical interferometers, atom interferometers split a matter wave into two parts, evolve the parts independently along different paths, and finally recombine the waves to form an interference pattern \cite{Cronin20019}. The interference pattern depends on the accumulated phase shift of the wave packets during the independent evolution, and the measured quantity is typically inferred from the shape and time evolution of the pattern. The sensitivity of interferometers increases with the accumulated phase shift, which again depends on the evolution time \cite{Debs2011}. However, the evolution time of a free falling atom cloud is limited by Earth's gravitational acceleration in most experimental setups, and considerable efforts are made to increase the duration, e.g.\ in fountain experiments \cite{Chiow2011}, drop towers \cite{Zoest2010,Muntinga2013}, parabolic flights \cite{Geiger2011,Barrett2016} and in space \cite{Becker2018}.

In this article, we employ magnetic levitation as a different method to extend the evolution time in earthbound laboratories. Magnetic levitation relies on the use of magnetic forces to cancel the gravitational acceleration and to levitate the particles in space. The method is well established for experiments with ultracold atoms \cite{Anderson1995, Han2001, Weber2003}, and its experimental implementation, i.e.\ using a pair of current-carrying coils, is significantly simpler and smaller than an atomic fountain apparatus or a drop-tower experiment. Here, we study the advantages and limitations of magnetic levitation for matter wave interferometry with the motional states of Bose-Einstein condensates (BECs), and we demonstrate that magnetic levitation can be employed to reach an expansion time of 1\,s, which is comparable to current drop-tower experiments \cite{Zoest2010}. Furthermore, we utilize magnetic levitation to create and to interferometrically measure micro-g acceleration in free expansion, and we show that the negligible center-of-mass motion of levitated atoms facilitates a direct study of phase-shifting elements in the interferometer paths.

Other interferometer schemes use external trapping potentials to prevent the gravitational acceleration by channelling the wave packets along magnetic \cite{Wang2005,Garcia2006} and optical \cite{McDonald2013,Marti2015} waveguides. External guiding and trapping potentials allow for equally long observation times \cite{Burke2008}, however, they introduce additional challenges. External potentials can cause spatially varying phase shifts and undesired excitations of the wave packets \cite{Burke2008,Marti2015}, which limit the measurement precision. Our levitation scheme avoids trapping potentials along the gravitational axis, and it facilitates a tuneable scattering length for future studies of interaction effects in atom interferometers.

This article is structured as follows:\ section\,\ref{sec:setup} provides an overview of our experimental setup, magnetic levitation scheme, and the use of a magnetic Feshbach resonance to control the interaction strength of cesium atoms. Section\,\ref{sec:interferometer} is used to illustrate the interferometer scheme, and in section\,\ref{sec:Measurements} we evaluate our measurement precision. Small changes to the magnetic levitation gradient allow us to create marginal accelerations of milli-g (section\,\ref{sec:MeasureMilli}) and micro-g (section\,\ref{sec:MeasureMicro}). An additional laser beam in one of the interferometer paths constitutes a phase-shifting element in section\,\ref{sec:MeasurePhase}. In section\,\ref{sec:curvature}, we measure features of the magnetic field distribution, such as the transversal curvature of the force field. Finally, using a combination of low interaction strength, low trapping frequencies, and magnetic levitation we demonstrate long expansion and observation times in section\,\ref{sec:expansion}.

\begin{figure}[t]
\centering
  \includegraphics[width=0.9\textwidth]{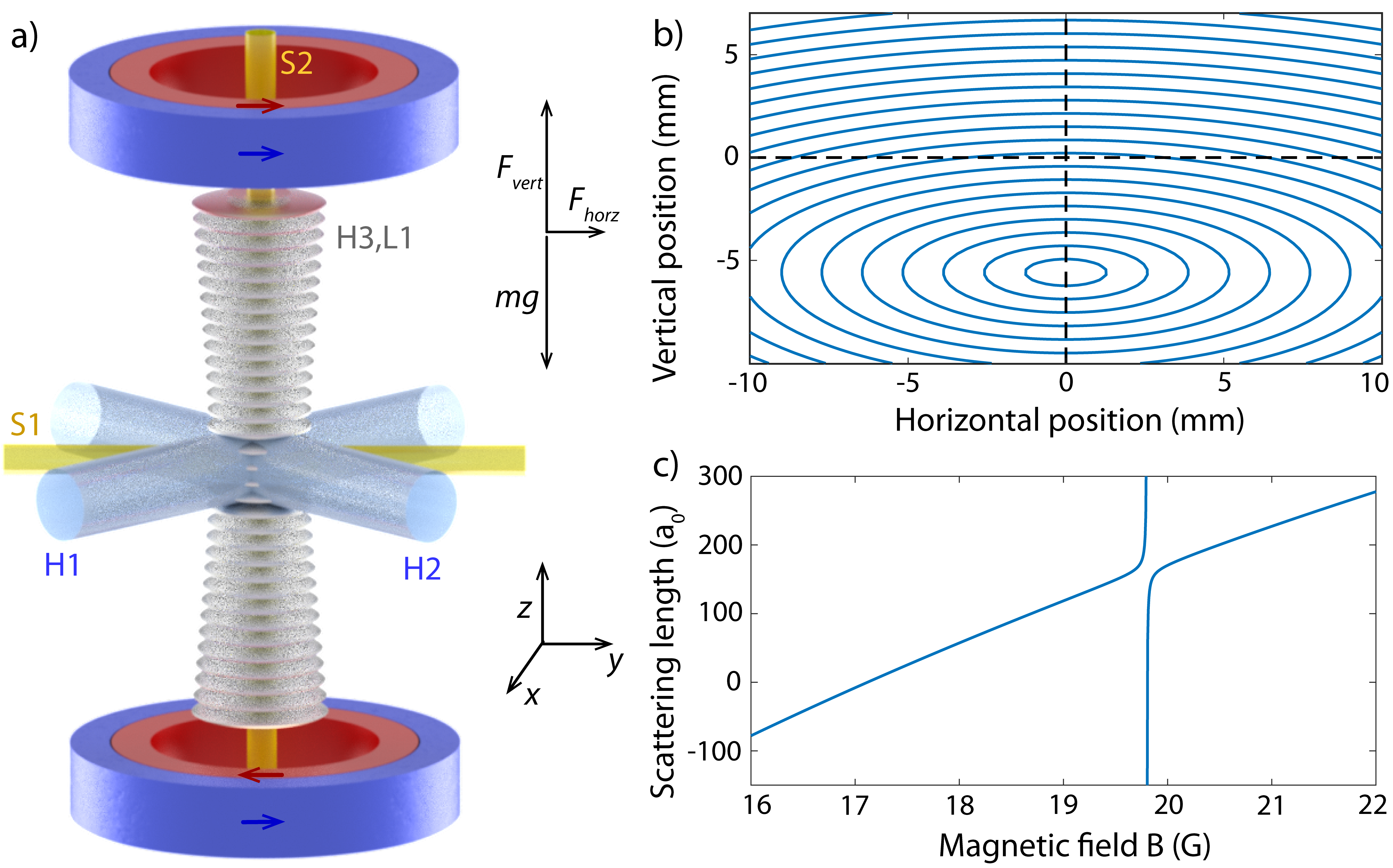}
  \caption{Experimental setup. a) Magnetic field coils to control $B_0$ (blue, outer coils) and $\partial_z B$ (red, inner coils). Laser beams with small beam waists (S1,S2) and large beam waists (H1,H2,H3) trap the atoms, and a lattice L1 is used to split the wave packet during the interferometer sequence. Top and bottom coils have an inner diameter of 12\,cm and a vertical separation of 6\,cm. b) Numeric simulation of the total magnetic field $|B(y,z)|$ for $\partial B/\partial z = 31.1$\,G/cm and $B_0 = 17.4$\,G, field lines indicate a magnetic field strength of 2\,G - 40\,G. c) Zero crossing of the scattering length at 17.1\,G due to a broad Feshbach resonance for cesium atoms.\label{Fig:setup}}
\end{figure}

\section{Magnetic levitation scheme and experimental apparatus\label{sec:setup}}

Our experimental apparatus is designed to independently control two parameters of the magnetic field. The magnetic field strength $B_0=|\mathbf{B}(x,y,z)|$, at the position of the atoms ($x\!=\!y\!=\!z\!=\!0$\,mm) is used to tune atomic interactions by means of a broad magnetic Feshbach resonance for cesium atoms in the strong-field-seeking Zeeman state $\ket{F=3, m_F=3}$. We reduce the effects of interaction by setting $B_0$ to 17.4\,G with an s-wave scattering length, $a$, of approximately 65\,$a_0$ during the interferometer sequences (Fig.\,\ref{Fig:setup}c), where $a_0$ is Bohr's radius. The second controlled parameter is the vertical gradient of the magnetic field, $\partial_z B$, which can be adjusted to exert a vertical pull on the atoms and cancel the gravitational acceleration. Due to the Zeeman effect, cesium atoms in the given state experience a vertical force that is proportional to the magnetic field gradient, $F_\mathrm{vert}= \frac{3}{4}\mu_B \partial_z B$. For a mass $m$ of a cesium atom, the levitation gradient can be calculated as $\partial_z B = 4mg/(3\mu_B) = 31.1\,$G/cm \cite{Weber2003,Kraemer2004}. Here, $\mu_B$ represents the Bohr magneton and $g$ the gravitational acceleration.

Our coil configuration is based on established designs \cite{Han2001,Weber2003,Kraemer2004}. It consists of two vertical coils above and below the atoms (inner diameter 12\,cm, separation 6\,cm), with 5 independently controllable sections. We generate $B_0$ and $\partial_z B$ by means of two vertical pairs of coil sections with co- and counter-propagating currents (outer and inner sections in Fig.\,\ref{Fig:setup}a). Pairs of shim coils on each axis at distances of approximately 20\,cm from the atoms allow for additional fine control of the magnetic field. Figure\,\ref{Fig:setup}b shows the total magnetic field strength $B(y,z)$ in the vertical plane as calculated by a numerical simulation of our coils with finite wire elements. The field can be approximated by a magnetic quadrupole field with a shifted minimum at a few millimetres below the atom cloud. Experimentally, we determine $B_0$ by microwave spectroscopy and we optimize the levitation gradient $\partial_z B$ by varying the levitation current $I_\mathrm{lev}$ and minimizing position drifts of a BEC during free levitated expansion. Additional effects due to horizontal field curvature and limitations of the levitations scheme for precision measurements are discussed in section\,\ref{sec:curvature}.

The matter waves of our interferometer are provided by Bose-Einstein condensates. In our setup, $2\times 10^9$ cesium atoms are loaded from a 2D+ magneto optical trap (MOT) into a 3D MOT within 3\,s. The atoms are cooled by degenerate Raman sideband cooling \cite{Kerman2000}, and then sequentially transferred into two pairs of crossed optical dipole traps, the first with wavelength 1070\,nm, total power 200\,W, waists 700\,$\mu$m, and the second with wavelength $\lambda = 1064.495(1)\,$nm, power 400\,mW, waists 90\,$\mu$m (labels S1, S2 in Fig.\,\ref{Fig:setup}a ). Bose-Einstein condensation is reached after 6\,s of evaporative cooling, and the density distribution of the atoms is detected by means of resonant absorption imaging after a variable time of levitated expansion and after 1 ms of unlevitated time-of-flight. One cooling cycle has a duration of 15\,s and it is similar to ref.~\cite{Kraemer2004}.

We generate BECs of $2.5\times 10^5$ atoms in the Zeeman sub-state $\ket{F = 3, m_F = 3}$ at a scattering length of $a = 210\,a_0$, trapped in the crossed laser beams S1, S2 with trap frequencies of $\omega_{\mathrm{x,y,z}}=2\pi\times(23.5,17.7,15.4)$\,Hz. To reduce interactions during the interferometric measurement, we tune the scattering length to 65\,$a_0$ and remove atoms by forced evaporation with a non-levitating magnetic field gradient. The BECs for the interferometer measurements in this work consist of approximately $8\times 10^4$ atoms with a thermal fraction below 5\%. Vibrational isolation and damping of the optical table is achieved by a pneumatic isolation system (Newport S-2000A).

\section{Interferometer scheme \label{sec:interferometer}}

\begin{figure}[t]
  \centering
  \includegraphics[width=0.9\textwidth]{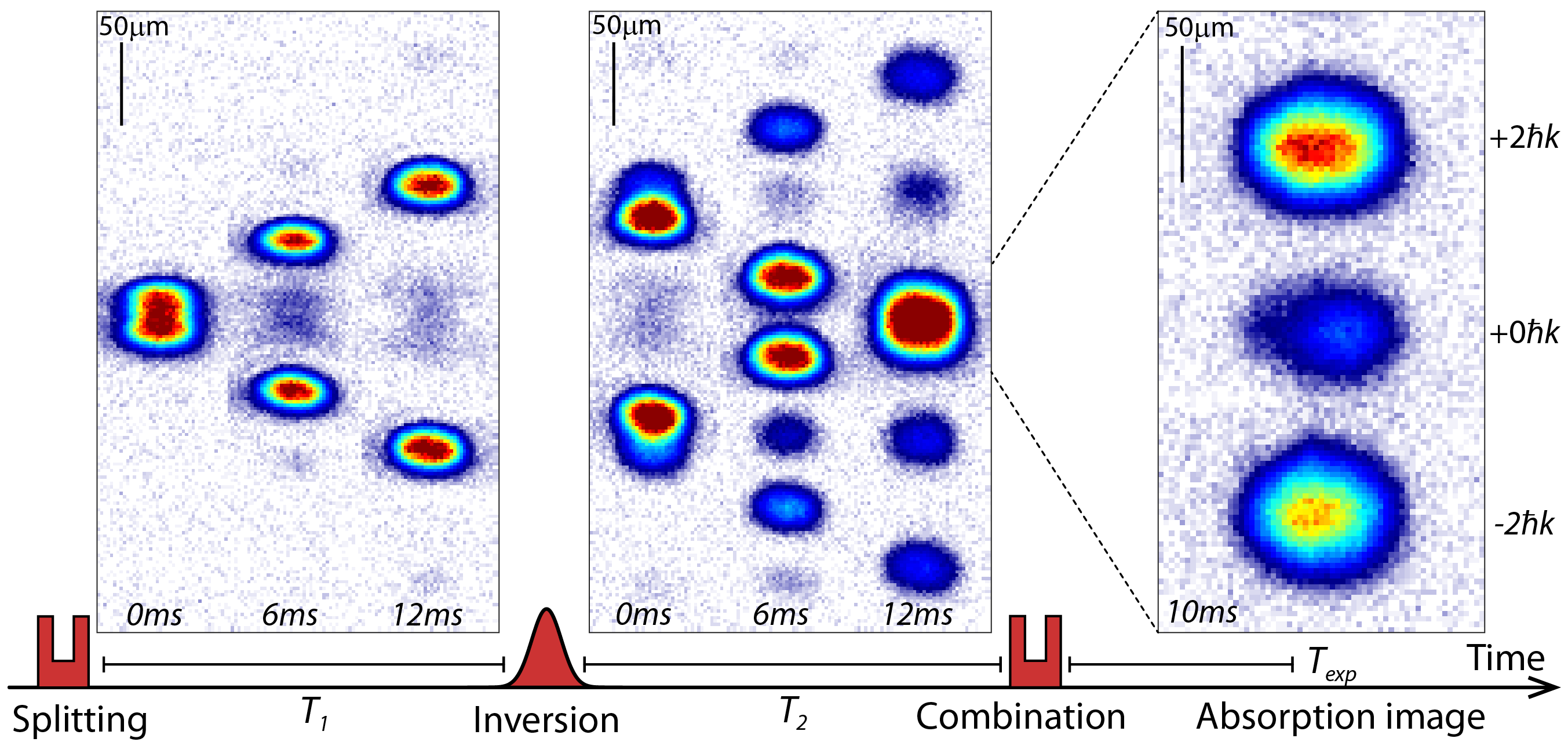}
  \caption[width=1\textwidth]{Interferometer scheme. Average of three absorption images of the matter waves after the splitting and the inversion pulses (left to right: $T_1=T_2=0$\,ms, 6\,ms, 12\,ms), and after the recombination pulse and an expansion time of 10\,ms. All images are taken after an additional time-of-flight of 1ms. \label{Fig:pulses}}
\end{figure}

We employ a Michelson interferometer scheme that is based on three Kapitza-Dirac pulses with a standing light wave (Fig.\,\ref{Fig:setup}a, beam L1) \cite{Gould1986}. The pulses change the motional states of the matter waves but leave the internal states of the atoms unchanged \cite{Rasel1995}. Our pulse sequence and the resulting motion of the matter wave packets are illustrated in Fig.\,\ref{Fig:pulses}. A first pulse splits the BEC into two wave packets with opposite momenta $\pm2\hbar k_L$. Here, $k_L=2\pi/\lambda$ is the wavenumber of the lattice beam and $\hbar$ is Planck's constant. The wave packets propagate freely for an evolution time $T_1$ until we apply a second pulse that inverts the direction of the wave packets and changes their momentum by $4\hbar k_L$. A third pulse is used after an evolution time $T_2$ to recombine the two wave packets. It is identical to the first pulse and generates three wave packets with momenta $p_0 = 0$, $p_{\pm} = \pm2\hbar k_L$. The relative population of the recombined wave packets depends on the acquired phase difference $\Delta \Phi$, resulting in a probability $P_0$ of finding an atom in the $p_0$ momentum mode
\begin{eqnarray}\label{eq:probability}
    P_0 = P_m + \frac{C}{2}\cos(\Delta \Phi).
\end{eqnarray} Here, $C$  is the interference contrast and $P_m$ is the offset of the interference signal. We determine $P_0$ from the ratio of atoms in the $p_0$ mode to the total atom number in all momentum modes.

Several factors can contribute to the phase difference $\Delta \Phi$. For falling wave packets with spatially homogeneous acceleration $a_c$, the phase difference is directly proportional to the center-of-mass displacement $\Delta z$ that was acquired during the total interferometer time $\Delta T=T_1+T_2+T_\mathrm{pulse}$. Here, $T_\mathrm{pulse}$ represents the total duration of the pulses. The total phase difference is given by \cite{Storey1994}
\begin{eqnarray}\label{eq:phase}
    \Delta \Phi = 2 k_L \Delta z + \Phi_0 =  2k_L\, \frac{1}{2} a_c (\Delta T)^2 + \Phi_0,
\end{eqnarray}
with a term $\Phi_0$ that accounts for additional phase shifts introduced during the initialization process, by noise such as lattice vibrations \cite{Barrett2016}, or by interactions (see section \ref{sec:MeasureMicro}).

The pulse sequence used in this experiment is based on previous work \cite{Wu2005,Wang2005,Robertson2017}. Our splitting and recombination pulses consist of three sub-pulses of lattice beam $L1$ with durations $60\,\mu$s, $110\,\mu$s, and $60\,\mu$s, and lattice intensities of 6.6$\,E_r$, 0.2$\,E_r$, and 6.6\,$E_r$. Here, $E_r=\hbar^2 k_L^2/(2m)$ is the recoil energy for cesium at a lattice wavelength of $1064$\,nm. Our inversion pulse has a Gaussian intensity distribution with a maximum of $17\,E_r$ and a $1/\mathrm{e}$-duration of $35\,\mu$s. The sub-pulse scheme allows us to reach a splitting efficiency of 96\% of the atoms in the $\pm 2\hbar k_L$ modes, and we speculate that the limit of the efficiency is given by the thermal component of our BEC. The efficiency of the inversion pulse is lower, 83\%, and residual atoms are clearly visible in Fig.\,\ref{Fig:pulses} in the 0 and $\pm 2 \hbar k_L$ modes. We suspect that this is due to the velocity selectivity of the inversion pulse and the velocity difference of the accelerated wave packets.

\section{Interferometric measurements \label{sec:Measurements}}

\begin{figure}[t!]
  \centering
  \includegraphics[width=1\textwidth]{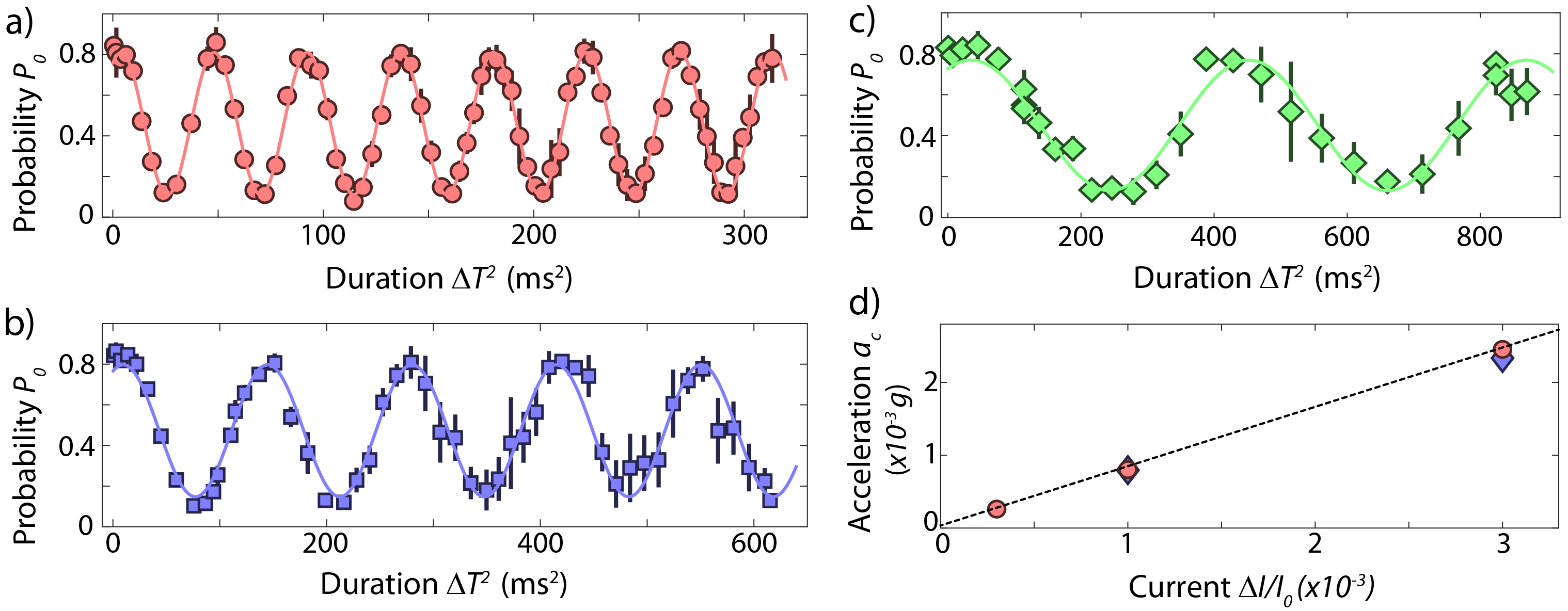}
  \caption{Interferometric measurement of milli-g accelerations. a-c) Probability of observing atoms in the $0\hbar k_L$ momentum mode for increasing duration $\Delta T$ and gradient coil currents $\Delta I/I_\textrm{lev}$ of a) 0.003, b) 0.001, c) 0.0003. Solid lines represent fits to the data points using eq.\,\ref{eq:probability} and \ref{eq:phase}. d) Comparison of the acceleration measurement with the interferometer scheme (red circles) and by the center-of-mass motion (blue diamonds). Error bars indicate one standard deviation of the data points.\label{Fig:gravity}}
\end{figure}

\subsection{Measuring milli-g acceleration\label{sec:MeasureMilli}}
Our magnetic levitation scheme allows us to apply small forces to the atoms by changing the levitation current $I_\mathrm{lev}$ in the vertical coils with counter-propagating currents. We use this approach to characterize our interferometer setup for non-zero accelerations. After the preparation of the BEC we increase the current $I$ in the coils, which create the magnetic field gradient, in 75\,ms to the ratios $I/I_\mathrm{lev}$ of $1.003, 1.001$, and  $1.0003$. The acceleration of the BEC is measured with our interferometer scheme. Figures\,\ref{Fig:gravity}\,a-c show the corresponding measurements of $P_0$ for varying evolution times $\Delta T^2$ with $T_1 = T_2$. As expected, we observe sinusoidal oscillations of $P_0$, which are fitted using eq.\,\ref{eq:probability} and \ref{eq:phase} (solid lines) to determine the accelerations $a_c$ (red circles, Fig.\,\ref{Fig:gravity}d).

An independent measurement of $a_c$, based on the free motion of the BEC, is provided for comparison. We measure the shift of the center-of-mass position for an expansion time $T_\mathrm{exp}$ of an untrapped BEC in our magnetic field gradients, $z(T_\mathrm{exp}) = 1/2\, a_c T_\mathrm{exp}^2$, with a fit parameter $a_c$ (blue diamonds, Fig.\,\ref{Fig:gravity}d). We find excellent agreement within two standard deviations between the two methods. However, the sensitivity of the free expansion measurement is limited by the observation time. Although our levitation scheme allows for very long observation times (section\,\ref{sec:expansion}), it also induces a horizontal dispersion of the BEC in free space, which will be discussed in section\,\ref{sec:curvature}. Here, we limit the observation time to 200\,ms, which allows us to measure the acceleration for $I/I_\mathrm{lev}=1.001, 1.003$, but not for 1.0003. The measurement results in Fig.\,\ref{Fig:gravity}d have relative uncertainties of approximately 4\% for the free expansion measurement and 0.5\% for the interferometric approach.

\subsection{Measuring micro-g acceleration\label{sec:MeasureMicro}}

\begin{figure}[t!]
  \centering
  \includegraphics[width=0.9\textwidth]{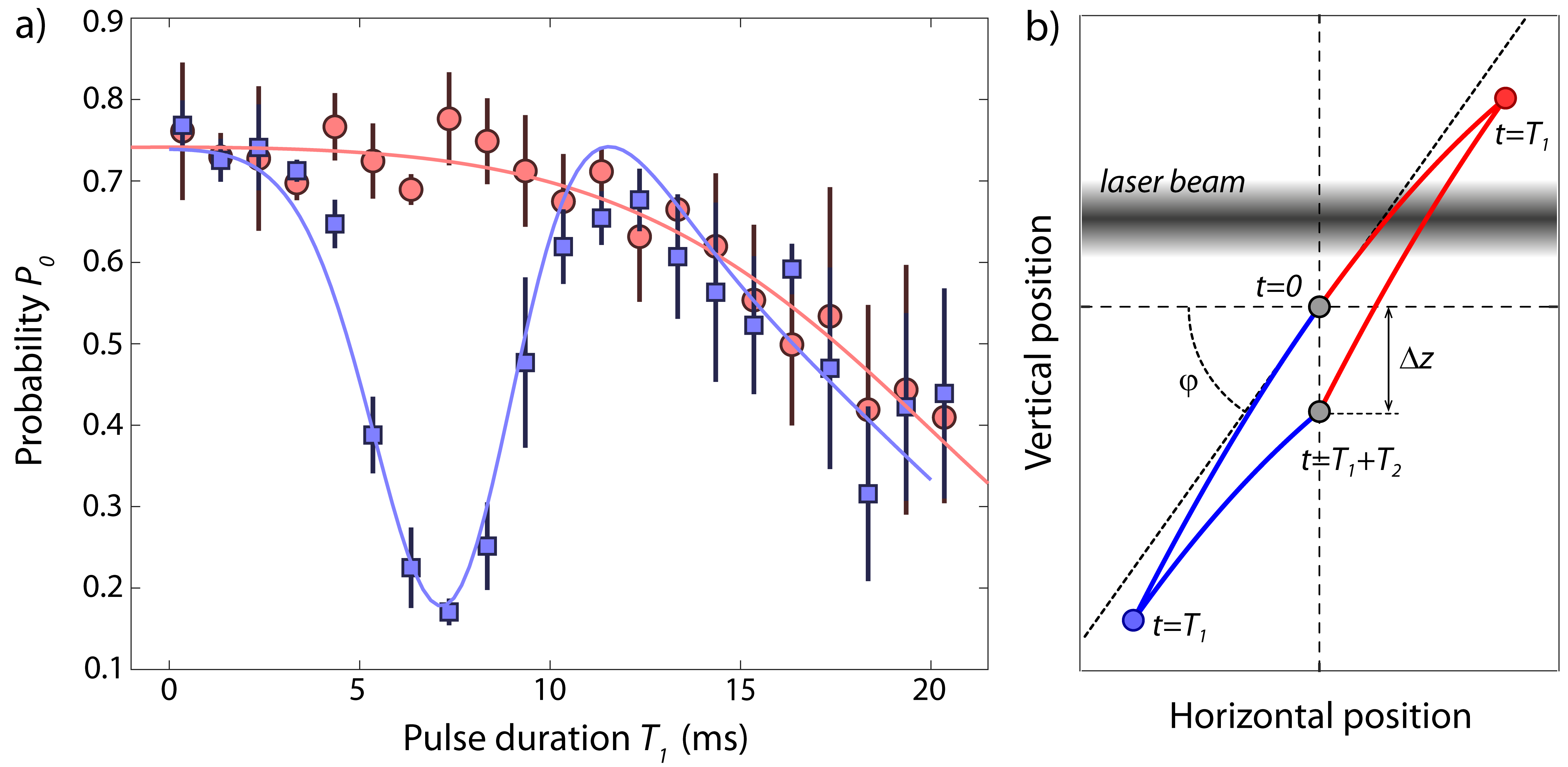}
  \caption{Interferometric measurement of micro-g accelerations and phase shifts due to a laser beam. a) Probability of observing atoms in the $0\hbar k_L$ momentum mode vs  $T_1$ for minimized acceleration of the atoms (red circles) and for an addition laser beam in the path of the upper wave packet (blue squares). Error bars indicate one standard deviation of the data points. b) Illustration of the position of the wave packets and the additional laser beam during the pulse sequence. Angles and axes are not to scale in the illustration. \ \label{Fig:zeroacceleration}}
\end{figure}

In a second measurement, we utilize the interferometer scheme to minimize the forces on the atoms. We vary the currents in our shim coils and $I_\mathrm{lev}$ with the goal to maximize the oscillation period of $P_0$ (red circles Fig.\,\ref{Fig:zeroacceleration}). For optimal current values, we observe a slow drop of the value of $P_0$ from approximately 0.75 to 0.45 over $\Delta T^2 \approx 1600$\,ms$^2$. This reduction is not necessarily caused by a residual acceleration of the wave packets, as it can also originate from dephasing mechanisms that are discussed in the next paragraph. However, fitting $P_0(t)$ with eq.\,\ref{eq:probability} provides an upper limit to the acceleration experienced by the atoms. We determine an upper limit for the acceleration of the atoms of $a_\textrm{c} = 70(10)\times 10^{-6}\,g$. Atomic fountain interferometers facilitate the measurement of significantly smaller differential accelerations and reach staggering precisions of the order $\Delta g/g \sim 10^{-10}$ \cite{Peters1999,Hardman2016,Asenbaum2017}. Our measurement, however, provides, to the best of our knowledge, the smallest absolute value for an acceleration that is measured directly with ultracold atom interferometry.

We estimate possible sources of measurement errors, fluctuations and dephasing mechanisms. Fluctuations of a homogeneous magnetic field will only slightly change the interaction strength of our BEC, but deviations of the magnetic field gradient can induce additional accelerations and alter the measurement result. In our setup, small deviations of the magnetic field gradient can occur as the wave packets move during an interferometer sequence away from the original position with optimized levitation. We estimate from our numerical magnetic field simulation that our coil design causes a relative increase of the field gradient of $2\times10^{-6}$ for a vertical position shift of 50\,$\mu$m. In addition, the quadratic Zeeman effect induces another deviation of the levitation force of $6\times10^{-6}$ for the same position shift. As a result, the upper and lower wave packets experience a position-dependent acceleration, which increases the separation of the wave packets before the inversion pulse, and which reduces the convergence after the inversion pulse. Similar to our measurements in section\,\ref{sec:curvature}, we would expect the final displacement of the wave packets to cause horizontal fringes in the absorption images, which we do not observe. As a result, we conclude that the vertical force gradients are negligible for the time scales of our interferometer.

In addition, the position-dependent magnetic field strength causes an almost linear change of the scattering length of approximately $\pm10\,a_0$ over 50\,$\mu$m (see also section \ref{sec:expansion}). As a result, the atoms in the upper wave packet experience a stronger interaction and faster phase evolution than atoms in the lower wave packet. Assuming constant densities and a linear change of the scattering length, we would expect the phase shift between the wave packets to increase with $\Delta T^2$, and it would be difficult to distinguish this effect from a phase evolution due to acceleration. However, in our setup the wave packets expand after release and the densities decrease strongly over a timescale of $1/\omega_{x,y,z} \approx 10$\,ms. The position-dependent scattering length would result in a change of the oscillation frequencies within 10-15\,ms in Fig.\,\ref{Fig:gravity}a-c, which we do not observe, and we conclude that the phase shift due to a position-dependent scattering length is below our sensitivity for this measurement.

Fluctuations of the acceleration of the BEC can be caused by time-dependent changes of $B_0$ and $\partial_z B$, either due to external magnetic fields or due to the finite stability of the currents in our coils. We determine a current reproducibility of $1.4\times10^{-6}$ by measuring the standard deviation of the current during the interferometer sequence over 60 consecutive cycles. For each cycle, the current measurement averages over 80\,ms. We believe that the current reproducibility will eventually set the limiting precision for our interferometric measurements with levitated atoms. While it is in principle possible to increase the current reproducibility by 1-2 orders of magnitude by improving our current regulation electronics, it would be very hard to reach the precision of atomic fountain experiments. Nonetheless, we believe that magnetic levitation schemes will provide a valuable technological addition for precision measurements with ultracold atoms. Reducing gravitational acceleration to micro-g effectively removes the center-of-mass motion of the atoms, and it allows for a direct measurement of phase-shifts due to additional elements in the interferometer path. We demonstrate this approach in the next section by adding a focused laser beam in the upper path of the interferometer and by measuring its position-dependent phase shift on the atoms.

\subsection{Detection of phase-shifting elements\label{sec:MeasurePhase}}
Compared to fountain experiments, the center-of-mass motion of our wave packets is contained within a small spatial region of a few hundreds of $\,\mu$m, and it is straightforward to add additional phase shifting elements in the path of the wave packets. As a result, it is possible to use the levitated interferometer scheme to analyze additional potentials for the atoms with high precision. We demonstrate this approach by adding a horizontal laser beam (wavelength 1064\,nm, waist $40\,\mu$m, power $29\,\mu$W) approximately $50\,\mu$m above the initial position of the atoms (Fig.\,\ref{Fig:zeroacceleration}b). This beam creates a Gaussian dipole potential with a depth of approximately 3\,nK, and it introduces between the upper and lower wave packets a differential phase shift, which can be detected by the interferometer. In addition to a measurement of the AC Stark shift of the light field as in reference \cite{Deissler2008}, our setup facilitates the study of the spatial dependence of the potential.

The effect of the laser beam on $P_0(t)$ is clearly visible in Fig.\,\ref{Fig:zeroacceleration}a when comparing the data sets with the beam (blue squares) and without the beam (red circles). For increasing duration $T_1$, the upper wave packet passes twice through the laser beam and it samples increasing spatial sections of the potential. We adjusted the power of the beam to create a single oscillation of the phase for a wave packet that fully transverses the beam, resulting in a minimum of $P_0(t)$ at an evolution time $T_1 = 7\,$ms in Fig.\,\ref{Fig:zeroacceleration}a.

Constant propagation velocities of the wave packets during the evolution times $T_1$ and $T_2$ make it easy to relate the time to the position of the atoms. We use a numerical model to integrate the phase shift of the upper wave packet in the dipole potential of the laser beam over the interferometer path $z(t)$ and include the unperturbed phase shift as measured in section\,\ref{sec:MeasureMicro}. Fitting the model parameters to our data set (blue line Fig.\ref{Fig:zeroacceleration}a), we determine a beam position of $45(1)\,\mu$m, a waist of $37(4)\,\mu$m and a beam power of 25(3)\,$\mu$W, which are in excellent agreement with the independently measured values.

Our model neglects the spatial extent of the wave packets and we determine the phase shift at the center-of-mass position, whereas our experimental sequence averages over local phase shifts within the upper matter wave packet. Local phase shifts result in density variations in the profiles of the momentum modes in our absorption images, but measuring the total atom number in the momentum modes provides only the average phase shift of the wave packet.

\section{Spatial curvature of the force field\label{sec:curvature}}

Our magnetic field configuration does not only provide a vertical magnetic field gradient to levitate the atoms, but it also generates a weak, horizontal anti-trapping potential. This potential is a result of the spatial curvature of our quadrupole-like distribution of the magnetic field (see Fig.\ref{Fig:setup}b). In this section, we demonstrate that the anti-trapping potential causes an additional interference pattern, which can be employed to measure the anti-trapping frequency or the angle between the lattice beam and the vertical field axis.

\begin{figure}[t!]
  \centering
  \includegraphics[width=1\textwidth]{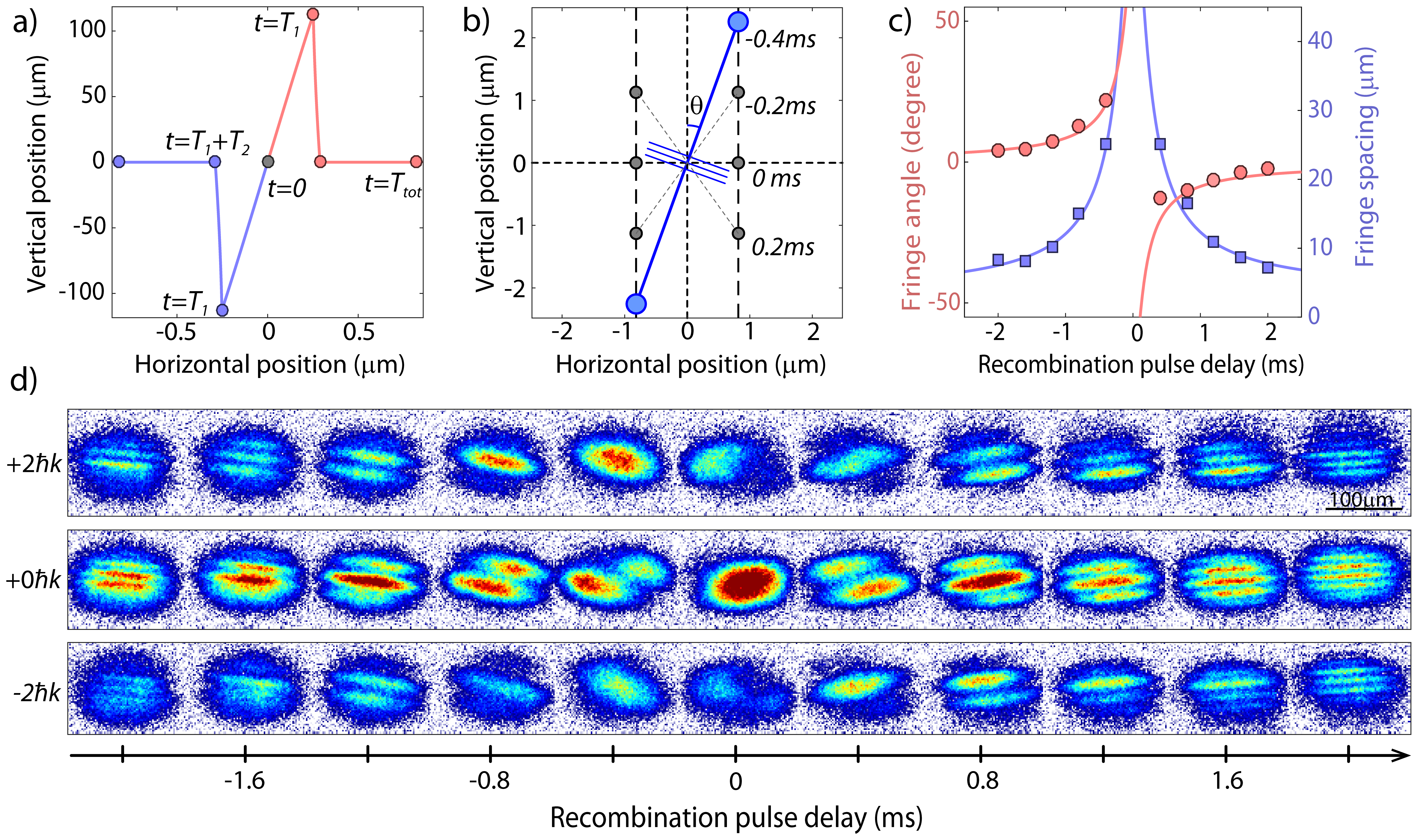}
  \caption{Effect of the force field curvature on the interference pattern. a) Calculated interferometer path of the center-of-mass positions of the levitated wave packets with $\delta t=0$\,ms. b) Center-of-mass positions of the two wave packets for $\delta t$ = -0.4\,ms (blue), -0.2\,ms, 0\,ms, +0.2\,ms (grey). Blue parallel lines indicate the orientation of the interference pattern. c) Fringe angles (red circles) and fringe spacings (blue squares) vs. the delay $\delta t$ of the recombination pulse, inferred from the data in d. Solid lines show our fit results for eq.~\ref{eq:displacement}. d) Absorption images for varying $\delta t$ between -2.0\,ms and 2.0\,ms in steps of 0.4\,ms. Common parameters are $\alpha = 2\pi\times 3.29$\,Hz, $\varphi=0.108^\circ$, $T_1=20\,$ms , $T_\mathrm{exp}$=30\,ms.\label{Fig:curvature} }
\end{figure}

Within the quadrupole approximation it is possible to derive simple equations for the magnetic field and for the forces along the dashed horizontal line in Fig.\,\ref{Fig:setup}b \cite{Weber2003,Kraemer2004,Sackett2006} \begin{equation}\label{eq:bfield}
  B_\mathrm{horz}(r) = B_0 + \frac{2}{9} \frac{m^2}{\mu_B}\frac{g^2}{B_0} r^2, \quad\qquad F_\mathrm{horz}(r) =  m \alpha^2 r \quad \mathrm{with} \quad \alpha = g \sqrt{\frac{m}{3\mu_B B_0}}.
\end{equation}
Here, $r=\sqrt{x^2+y^2}$ is the horizontal displacement of the atoms from the origin. The quadratic scaling of $B_\mathrm{horz}(r)$ with $r$ results in a weak, outwards-directed force in the horizontal plane. This anti-trapping effect can be associated with frequency $\alpha$, and it causes a weak, position-dependent acceleration with a time-dependent horizontal position $r(t)$ and horizontal velocity $v_r(t)$ \cite{Herbig2003}: \begin{eqnarray}
    r(t) &= r(0) \cosh(\alpha t) + \alpha^{-1} v_r(0) \sinh(\alpha t)  \nonumber\label{eq:motion}\\
    v_r(t) &= v_r(0) \cosh(\alpha t) + \alpha r_0  \sinh(\alpha t)  \nonumber \\
    z(t) &= v_z(0) t + z(0).
\end{eqnarray}
For this calculation we assume perfect levitation and linear vertical motion $z(t)$ during the interferometer sequence.

In an experimental setup there will always be a small angle $\varphi$ between the lattice beam L1 and the vertical axis of the magnetic field, and a splitting pulse will always imprint a small velocity component $v_r(0)=(\hbar k_L/m) \sin(\varphi)$ along the horizontal direction. Consequently, a small horizontal displacement due to $v_r(0)$ results in an outwards-directed force on the wave packets in the anti-trapping potential, and in a finite horizontal displacement at the end of the interferometer sequence as illustrated in Fig.\,\ref{Fig:curvature}a. The horizontal distance between the wave packets is typically two orders of magnitude smaller than the vertical displacement during the interferometer sequence, and both distances become comparable only in the proximity of the recombination pulse and during the expansion time. We illustrate the positions of the wave packets in Fig.\,\ref{Fig:curvature}b for small delay times of the recombination pulse $\delta t= T_2-T_1$ with $T_1=20\,$ms. Depending on $\delta t$, the orientation of the blue line connecting the wave packets changes from almost vertical for $\delta t = \pm 0.4$\,ms to horizontal for $\delta t = 0$\,ms. We define an angle $\theta$, which is chosen to be positive clockwise and in the interval $[-90^{\circ}, 90^{\circ}]$, to indicate the orientation of the line, and we define $d(\delta t)$ to be the distance between the two wave packets.

In analogy to Young's double slit experiment \cite{Andrews1997,Muntinga2013}, the interference pattern of two wave packets at distance $d(\delta t)$ shows a fringe spacing $d_\mathrm{F}$ of
\begin{eqnarray} \label{eq:fringespacing}
    d_\mathrm{F}=\pi\hbar t/(m d)+d_0.
\end{eqnarray}
Here, $t$ is the total duration of the interferometer sequence with $t=T_1+T_2+T_\mathrm{pulse}-\delta t + T_\mathrm{exp}$, and $d_0\ge 0$ is a constant phase shift that depends on the initial conditions such as the density distribution \cite{Wallis1997,Roehrl1997,Simsarian2000}.
In our absorption images of the interfering wave packets for constant times $T_1,T_\mathrm{exp}$ and varying delay $\delta t$ (Fig.\,\ref{Fig:curvature}d), interference fringes with varying separation $d_\mathrm{F}$ and angle $\theta$ are clearly visible for all momentum modes $p_0, p_{\pm}$.

From the evolution of the fringes as a function of time delay $\delta t$, we infer properties of the curvature $\alpha$ and the angle $\varphi$. We simultaneously fit the fringe spacing in eq.\,\ref{eq:fringespacing} and the fringe angle $\theta$ with $\label{eq:fringeangle} \theta(\delta t)=\arctan(z(\delta t)/r(\delta t))$. Here $z(\delta t)$ and $r(\delta t)$ are the vertical and horizontal positions of the wave packets for varying $\delta t$. We integrate the center-of-mass motion of the wave packets in eq.\,\ref{eq:motion} with starting conditions $r(0)\!=z(0)\!=\!0$ over all steps of the interferometer sequence to determine $z(\delta t)$ and $r(\delta t)$
\begin{align}\label{eq:displacement}
  z(\delta t) = & -v_z(0) \delta t \nonumber \\
  r(\delta t) = & \frac{v_r(0)}{\alpha} \cosh(\alpha T_\mathrm{exp})\,  \bigg[ \sinh(\alpha T_1)\cosh(\alpha (T_1+\delta t))+   \bigg. \nonumber   \\
        &                          \qquad \qquad\qquad\qquad   \bigg.   (\cosh(\alpha T_1)-1)\sinh(\alpha(T_1+\delta t)) \bigg]  + \nonumber \\
        &  \frac{v_r(0)}{\alpha} \sinh(\alpha T_\mathrm{exp})\, \bigg( \bigg[ \sinh(\alpha T_1)\sinh(\alpha(T_1+\delta t))+ \bigg.\bigg. \nonumber \\
        &                           \qquad \qquad\qquad\qquad \bigg.\bigg.   (\cosh(\alpha T_1)-1)\cosh(\alpha(T_1+\delta t))\, \bigg] +1 \bigg)    .
\end{align}
Equations \ref{eq:displacement} contain two free parameters, the anti-trapping frequency $\alpha$ and the lattice angle $\varphi$, which can both be used to fit our data points in Fig.\,\ref{Fig:curvature}c. We choose to constrain $\alpha$ and vary $\varphi$ during the fitting procedure, as it is experimentally difficult to determine the laser beam angle with milliradian precision, and we independently measured $\alpha$ by observing center-of-mass oscillations of BECs in optical dipole traps. The fit results, represented by solid lines in Fig.\,\ref{Fig:curvature}c, show good agreement with our data points, and we measure a lattice angle of $\varphi=0.108(7)^{\circ}$ for $\alpha=2\pi\times 3.29(5)\,$Hz.

Note that $\alpha$ scales with $1/\sqrt{B_0}$ in eq.~\ref{eq:bfield}, and we can use larger values for $B_0$ to reduce the anti-trapping effect, e.g.\ by tuning the interaction strength with a broad magnetic Feshbach resonance at 800\,G \cite{Berninger2013}. However, it will be difficult to reduce $\alpha$ significantly due to its square-root dependence on $B_0$. Instead, it is easier to compensate the anti-trapping effect with an additional dipole trap, as demonstrated in the next section.

\section{Long expansion times\label{sec:expansion}}

\begin{figure}[t!]
  \centering
  \includegraphics[width=1\textwidth]{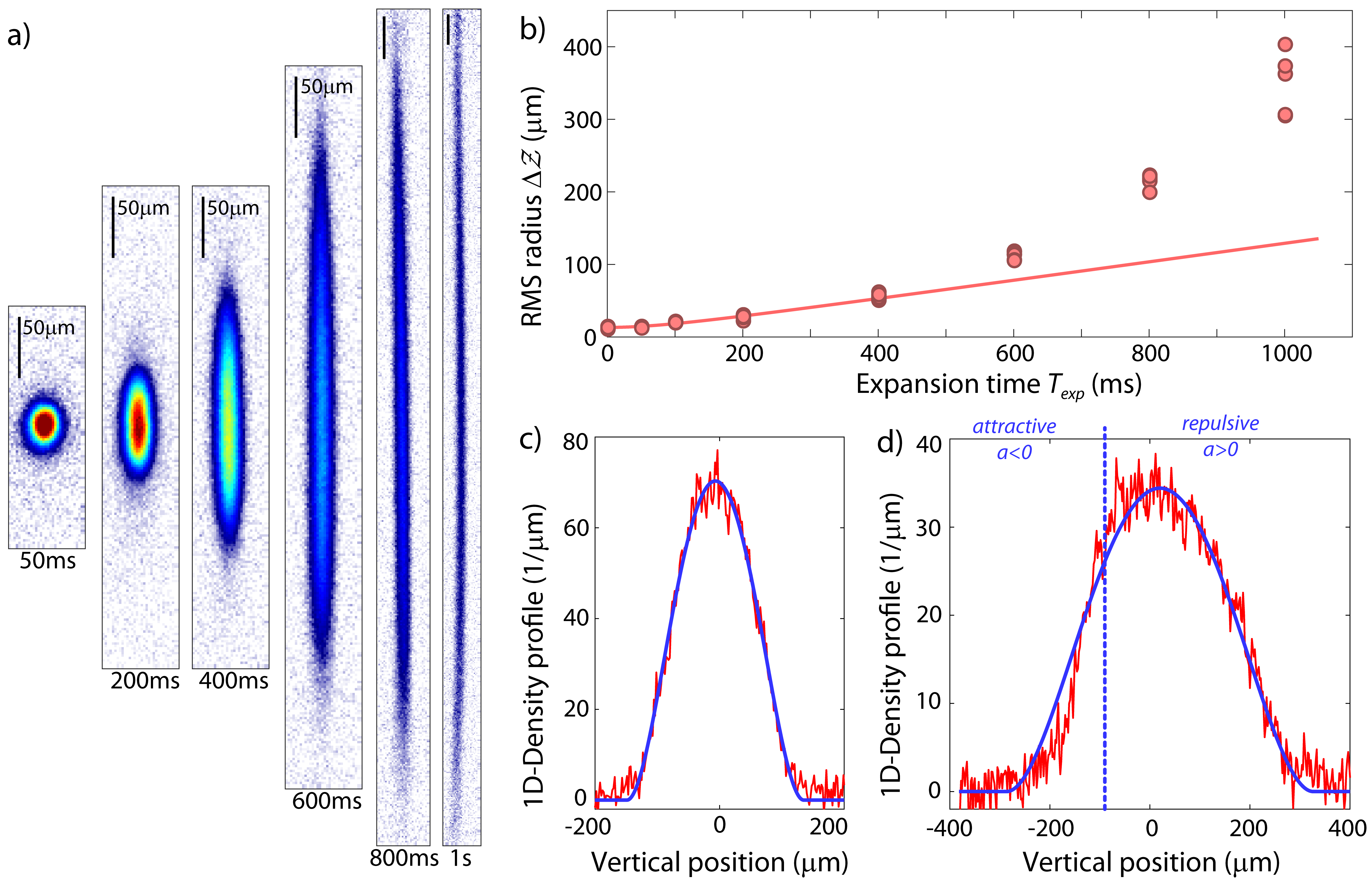}
  \caption{Long expansion times. a) Average of 6-8 absorption images for each expansion time: $T_\mathrm{exp} = 50,200,400,600,800,1000\,$ms. Note that the scaling of the images changes as indicated by the $50\,\mu$m scale bar in each picture. b) RMS-widths of the integrated 1D-density distribution vs. expansion time. c,d) 1D-density profiles and fits (blue lines) for expansion times of c) 400\,ms and d) 600\,ms. \label{Fig:expansion}}
\end{figure}

The sensitivity of an interferometric measurement increases with the evolution time of the wave packets \cite{Zoest2010}, but even without the implementation of an interferometer scheme, long observation times of an expanding BEC facilitate a sensitive acceleration measurement. In this section, we demonstrate that magnetic levitation allows us to extend the expansion time of a BEC to 1\,s, and we evaluate advantages and limitations of this scheme for precision measurements.

Typical expansion times for falling BECs are on the order of tens of milliseconds, often limited by the detection area of the imaging system, by the gravitational acceleration and by the expansion velocity of the gas. Usually, the expansion velocity of a quantum gas is not caused by the temperature of the gas but by repulsive interaction during the initial spreading. The current record for long observation times under milli-g acceleration is 1\,s \cite{Zoest2010} with an expansion energy of $9\,$nK. The experiment was performed in a drop tower, and ballistic expansion was observed over approximately $500$\,ms, limited by stray magnetic fields.

In our experiment, we can reduce the interaction energy of the BEC by tuning the scattering length close to $0\,a_0$ by means of a magnetic Feshbach resonance (Fig.\,\ref{Fig:setup}c). Further reduction of the expansion energy has been demonstrated by rapidly changing the scattering length from a positive value to $0\,a_0$ during trap release \cite{Kraemer2004}, but we refrain from using this trick to avoid excitations of the BEC during release. Our horizontal magnetic field curvature (section\,\ref{sec:curvature}) introduces another limitation. During long observation times, the BEC expands horizontally into regions with a lower magnetic field gradient, causing a position-dependent sag of the density profile. In addition, small fluctuations of the horizontal magnetic field can break the symmetry and introduce slow horizontal drifts. We suppress both effects by keeping a vertical laser beam (H3 in Fig.\,\ref{Fig:setup}a) on during the expansion time, thus observing free expansion only in the vertical direction.

In detail, we reduce the trap frequency by slowly transferring the atoms from a crossed dipole trap of beams S1,S2 to a crossed dipole trap of beams H1, H2, and H3 with final trap frequencies of $\omega_{x,y,z}=2\pi\times(3.2, 3.4, 2.1)\,$Hz, a scattering length of $15\,a_0$ and atom numbers of approximately $1.1\times10^4$. Excitations of the BEC during the transfer are suppressed by smooth changes of the potential with a total transfer duration of 4\,s. After an additional settling time of 1\,s we switch off the horizontal beams H1 and H2 and study the expansion of the BEC in the vertical beam H3. The vertical trapping frequency of the laser beam H3 is approximately 25\,mHz, and the resulting fractional reduction of the expansion width after 1\,s is $6\times10^{-4}$, which is far below our measurement sensitivity for the width of the BEC.

The expansion of the BEC in the vertical direction is clearly visible on absorption images (Fig.\,\ref{Fig:expansion}a) for expansion times 0 to 1000\,ms, and horizontally-integrated 1D density profiles for expansion times of $400$\,ms and $600$\,ms are given in Fig.~\ref{Fig:expansion}c and d. Although the trapped BEC is initially only weakly confined with almost symmetric trap frequencies, it changes dimensionality during the expansion process in the vertical beam. The density of the BEC decreases strongly during the vertical expansion, and the chemical potential becomes smaller than the transversal harmonic oscillator energy $\hbar \omega_{x,y}$ as required for a quasi-1D description \cite{Petrov2004}. As a result, we do not expect a shape-preserving spreading of the density distribution for a 1D expansion because the BEC passes through various interaction regimes as its density decreases  \cite{Ohberg2002,Pedri2003}. For illustration, we show a fit to the upper 80\% of the 1D-density profiles $n(z)$ for the ''3D cigar''-regime \cite{Menottti2002} (Fig.\,\ref{Fig:expansion}c), but we refrain from a complete analysis of the density profiles, which is beyond the scope of this article. Instead, we quantify the width of the expanding BEC with the root-mean-square (RMS) radius $\Delta \mathcal{Z} = \left( \frac{1}{N} \int n(z) (z-\bar{z})^2\right)^{1/2}$ to provide an estimate of the expansion velocity (red circles Fig.\,\ref{Fig:expansion}b). Here, $\bar{z}$  is the center-of-mass position of the atoms. We observe an initial interaction driven expansion and a ballistic flight for $T_\mathrm{exp}\le400\,$ms with an RMS expansion velocity of $v_\mathrm{rms} = 0.128(5)$\,mm/s and a corresponding kinetic energy of $m v_\mathrm{rms}^2/2 = 1/2\, k_B\times 260(20)\,$pK. We note that this is the expansion energy of the BEC component, but not the initial temperature of the trapped quantum gas.

Similar to reference \cite{Zoest2010}, we find an accelerated expansion for longer expansion times, $T_\mathrm{exp}>500\,$ms. We expect that the dominant source of the accelerated expansion is the curvature of our levitation gradient due to the quadratic Zeeman effect and due to our coil design, as discussed in section~\ref{sec:MeasureMicro}. However, the density profiles of the atoms on the absorption images indicate two other contributions. We observe small radial oscillations for long expansion times after release from the trap in the guiding beam H3 (see image $T_\mathrm{exp}=1$\,s in Fig.\,\ref{Fig:expansion}a). Those oscillations can couple to the vertical motion or they can distort the radially integrated density distribution. In addition, we observe asymmetric 1D density profiles $n(z)$ for $T_\mathrm{exp}>500\,$ms (Fig.\,\ref{Fig:expansion}d). The profiles show a slower expansion velocity for the lower part of the cloud than for the upper part. We assume that this effect is caused by the position-dependent scattering length due to our magnetic field gradient. The zero-crossing of $a$ is indicated in Fig.\,\ref{Fig:expansion}d by a dashed blue line. This asymmetric expansion of a BEC with position-dependent scattering length requires further investigation that is beyond the scope of this article. We find small position fluctuations for long expansion times $T_\mathrm{exp}>400\,$ms of the BEC due to the finite current stability for the magnetic field gradient (section\,\ref{sec:MeasureMicro}). For illustration, we re-centered the center-of-mass position in the absorption images for the averaging process in Fig.\,\ref{Fig:expansion}a, but all other data in Fig.\,\ref{Fig:expansion}b-c results from the analysis of individual absorption images.

\section{Conclusion}

In conclusion, we experimentally studied the benefits and challenges of the use of magnetic levitation schemes for interferometric precision measurements with ultracold atoms. We employed a Michelson-type interferometer setup with BECs with tuneable interaction and magnetic levitation to demonstrate absolute acceleration measurements in the micro-g regime and we used the negligible center-of-mass motion of levitated atoms to study the position-dependent phase shift of the dipole potential of a focused laser beam. Moreover, we demonstrated expansion times of 1\,s for a BEC, which is comparable to current drop tower experiments, and we used an extrapolation method for the fringe patterns to study the curvature of a force field that acts perpendicularly to our interferometer setup.

In our setup, limitations of the sensitivity arise from magnetic field fluctuations due to the current regulation, and from position-dependent interactions and magnetic field gradients. Although the sensitivity in our setup is significantly lower than the sensitivity of atomic fountain experiments, we believe that levitation schemes provide interesting features with the prospect of technical applications. Cancelling gravitational acceleration offers the possibility to combine long observation times with compact interferometer setups. Interesting applications are the measurement of local variations of electric and magnetic fields, and of mean field effects due to atomic interactions.

\section*{Acknowledgements}
The authors would like to thank E. Riis and P.F. Griffin for helpful discussions. We acknowledge financial support by the EU through the Collaborative Project QuProCS (Grant Agreement 641277). AdC acknowledges financial support by EPSRC and SFC via the International Max-Planck Partnership. The work was also supported in part by the EPSRC Programme Grant DesOEQ (Grant No EP/P009565/1).


\section*{References}

\begin{thebibliography}{9}
\bibitem{Gustavson1997} Gustavson T L, Bouyer P and  Kasevich M A 1997 \textit{Phys. Rev. Lett.} \textbf{78} 2046.
\bibitem{Dutta2016} Dutta I,  Savoie D,  Fang B, Venon B, Garrido Alzar C L, Geiger R and Landragin A 2016 \textit{Phys. Rev. Lett.} \textbf{116} 183003.
\bibitem{Peters1999} Peters A, Chung K Y and Chu S 1999 \textit{Nature} \textbf{400} 849.
\bibitem{Hardman2016} Hardman  K S, Everitt P J, McDonald G D, Manju P, Wigley P B, Sooriyabandara M A, Kuhn C C N,  Debs J E, Close J D and  Robins N P 2016 \textit{Phys. Rev. Lett.} \textbf{117} 138501.
\bibitem{Bouchendira2011} Bouchendira  R, Clad\'{e} P, Guellati-Kh\'{e}lifa S, Nez F and Biraben F 2011 \textit{Phys. Rev. Lett.} \textbf{106} 080801.
\bibitem{Rosi2014} Rosi G, Sorrentino F, Cacciapuoti L, Prevedelli M and Tino G M 2014 \textit{Nature} \textbf{510} 518.
\bibitem{Prevedelli2014} Prevedelli M, Cacciapuoti L, Rosi G, Sorrentino F and Tino G M 2014 \textit{Phil. Trans. R. Soc. A} \textbf{372} 20140030.
\bibitem{Hamilton2015} Hamilton P, Jaffe M, Haslinger P, Simmons Q, M\"uller H and Khoury J 2015 \textit{Science} \textbf{349} 849.
\bibitem{Cronin20019} Cronin A D, Schmiedmayer J and Pritchard D E 2009 \textit{Rev. Mod. Phys.} \textbf{81} 1051.
\bibitem{Debs2011} Debs J E, Altin P A, Barter T H, D\"oring  D, Dennis G R, McDonald G, Anderson R P, Close J D and Robins N P 2011 \textit{Phys. Rev. A} \textbf{84} 033610.
\bibitem{Chiow2011} Chiow S-W, Kovachy T, Chien H-C and Kasevich M A 2011 \textit{Phys. Rev. Lett.} \textbf{107} 130403.
\bibitem{Zoest2010} van Zoest T \textit{et al} 2010 \textit{Science} \textbf{329} 1540.
\bibitem{Muntinga2013} M\"untinga H  2013 \textit{Phys. Rev. Lett.} \textbf{110} 093602.
\bibitem{Geiger2011} Geiger R \textit{et al} 2011 \textit{Nat. Commun.} \textbf{2} 474.
\bibitem{Barrett2016} Barrett B, Antoni-Micollier L, Chichet L, Battelier B, L\'{e}v\`{e}que T, Landragin A and Bouyer P 2016 \textit{Nat. Commun.} \textbf{7} 13786.
%\bibitem{Cho2017}  A. Cho, Science \textbf{357} (6355), 986-989.
\bibitem{Becker2018} Becker D \textit{et al} 2018 \textit{Nature} \textbf{562} 391.
\bibitem{Anderson1995} Anderson M H, Ensher J R, Matthews M R, Wieman C E and Cornell E A 1995 \textit{Science} \textbf{269} 198.
\bibitem{Han2001} Han D J, DePue M T and  Weiss D S 2001 \textit{Phys. Rev. A} \textbf{63} 023405.
\bibitem{Weber2003} Weber T, Herbig J, Mark M, N\"agerl H-C and Grimm R 2003 \textit{Science} \textbf{299} 232.
%\bibitem{Jamison2011} Alan O. Jamison, J. Nathan Kutz, and Subhadeep Gupta,  Phys. Rev. A, \textbf{84}, 043643 , (2011).
\bibitem{Wang2005} Wang Y-J, Anderson D Z, Bright V M, Cornell E A, Diot Q, Kishimoto T, Prentiss M, Saravanan R A, Segal S R and Wu S 2005 \textit{Phys. Rev. Lett.} \textbf{94} 090405.
\bibitem{Garcia2006} Garcia O, Deissler B, Hughes K J, Reeves J M, and Sackett C A 2006 \textit{Phys. Rev. A} \textbf{74} 031601(R).
\bibitem{McDonald2013} McDonald G D, Kuhn C C N, Bennetts S, Debs J E, Hardman K S, Johnsson M, Close J D, and Robins N P, 2013 \textit{Phys. Rev. A} \textbf{88}, 053620.
\bibitem{Marti2015} Marti G E, Olf R, and Stamper-Kurn D M 2015 \textit{Phys. Rev. A} \textbf{91} 013602.
\bibitem{Burke2008} Burke J H T, Deissler B, Hughes K J, and Sackett C A 2008 \textit{Phys. Rev. A} \textbf{78} 023619.
\bibitem{Kraemer2004} Kraemer T, Herbig  J, Mark M, Weber T, Chin C, N\"agerl H-C and Grimm R 2004 \textit{Appl. Phys. B} \textbf{79} 1013.
\bibitem{Kerman2000}  Kerman A J, Vuletic V, Chin C, and Chu S 2000 \textit{Phys. Rev. Lett.} \textbf{84} 439.
\bibitem{Gould1986} Gould P L, Ruff G A and Pritchard D E 1986 \textit{Phys. Rev. Lett.} \textbf{56} 827.
\bibitem{Rasel1995}Rasel E M, Oberthaler M K, Batelaan H, Schmiedmayer J and Zeilinger A 1995 \textit{Phys. Rev. Lett.} \textbf{75} 2633.
\bibitem{Storey1994} Storey P and Cohen-Tannoudji C 1994 \textit{J. Phys. II} \textbf{4} 1999.
\bibitem{Wu2005} Wu S, Wang Y-J, Diot Q and Prentiss M 2005 \textit{Phys. Rev. A} \textbf{71} 043602.
\bibitem{Robertson2017} Robertson B I, MacKellar A R, Halket J, Gribbon A, Pritchard J D, Arnold A S, Riis E and Griffin P F 2017 \textit{Phys. Rev. A} \textbf{96} 053622.
\bibitem{Asenbaum2017} Asenbaum P, Overstreet C, Kovachy T, Brown D D, Hogan J M and Kasevich M A 2017 \textit{Phys. Rev. Lett.} \textbf{118} 183602.
\bibitem{Deissler2008} Deissler B, Hughes K J, Burke J H T and Sackett C A 2008 \textit{Phys. Rev. A} \textbf{77} 031604.
\bibitem{Sackett2006} Sackett C A 2006 \textit{Phys. Rev. A} \textbf{73} 013626.
\bibitem{Herbig2003} Herbig J, Kraemer T, Mark M, Weber T, Chin C, N\"agerl H-C and Grimm R 2003 \textit{Science} \textbf{301} 1510.
\bibitem{Andrews1997} Andrews M R, Townsend C G, Miesner H-J, Durfee D S, Kurn D M and Ketterle W 1997 \textit{Science} \textbf{275} 637.
\bibitem{Wallis1997}Wallis H, R\"ohrl A, Naraschewski M and Schenzle A 1997 \textit{Phys. Rev. A.} \textbf{55} 2109.
\bibitem{Roehrl1997} R\"ohrl A, Naraschewski M, Schenzle A and Wallis H 1997 \textit{Phys. Rev. Lett.} \textbf{78} 4143.
\bibitem{Simsarian2000} Simsarian J E, Denschlag J, Edwards M, Clark C W, Deng L, Hagley E W, Helmerson K, Rolston S L and Phillips W D 2000 \textit{Phys. Rev. Lett.} \textbf{85} 2040.
\bibitem{Berninger2013} Berninger M, Zenesini A, Huang B, Harm W, N\"agerl H-C, Ferlaino F, Grimm R, Julienne P S and Hutson J M 2013 \textit{Phys. Rev. A} \textbf{87} 032517.
\bibitem{Petrov2004} Petrov D S, Gangardt D M and Shlyapnikov G V 2004 \textit{J. Phys. IV}  \textbf{116} 5.
\bibitem{Ohberg2002} \"Ohberg P and Santos L 2002 \textit{Phys. Rev. Lett.} \textbf{89} 240402.
\bibitem{Pedri2003} Pedri P, Santos L, \"Ohberg P and Stringari S 2003 \textit{Phys. Rev. A} \textbf{68} 043601.
\bibitem{Menottti2002} Menotti C and Stringari S 2002 \textit{Phys. Rev. A} \textbf{66} 043610.
\end{thebibliography}
\end{document}